\documentclass[final,3p,times,onecolumn]{elsarticle}

\usepackage{graphicx}
\usepackage{epsfig}
\usepackage{amsmath,amsfonts, amsthm, mathrsfs, amssymb, bbm}
\usepackage{url}
\usepackage{mathtools}
\usepackage{commath}
\usepackage[sc,osf]{mathpazo}
\usepackage[noend, ruled,vlined,noresetcount]{algorithm2e}
\usepackage{natbib}
\usepackage{times}
\usepackage{subfigure}
\usepackage{lineno}
\usepackage{caption}
\usepackage{amssymb}
\usepackage{multirow}
\allowdisplaybreaks
\journal{arXiv}

\begin{document}

\begin{frontmatter}

\title{Too little, too late -- a dynamical systems model for gun-related violence and intervention}

\author[DMath,DMed]{Feng Fu\corref{ff}}
\ead{fufeng@gmail.com}
\author[DMath,SFI]{Daniel N. Rockmore}
\ead{Daniel.N.Rockmore@dartmouth.edu}

\address[DMath]{Department of Mathematics, Dartmouth College, Hanover, NH 03755, USA}

\address[DMed]{Department of Biomedical Data Science, Geisel School of Medicine at Dartmouth, Lebanon, NH 03756, USA}

\address[SFI]{The Santa Fe Institute, 1399 Hyde Park Road, Santa Fe, NM 87501, USA}

\cortext[ff]{Corresponding author at: 27 N. Main Street, 6188 Kemeny Hall, Department of Mathematics, Dartmouth College, Hanover, NH 03755, USA. Tel.: +1 (603) 646 2293}

\begin{abstract}
In the United States the laws regulating the carrying of firearms in public vary state-to-state. In a highly publicized event, the Governor of New Mexico recently issued an emergency order temporarily banning the carrying of firearms in some areas -- and thus  rescinding the right-to-carry law in New Mexico -- after a spate of gun violence, citing a statistical threshold of general societal violence under which right-to-carry laws should be superseded. In this paper we frame this policy intervention as a dynamical systems model that measures the incidence of gun violence as a function of gun prevalence. We show that the Governor's emergency order -- fittingly issued as an emergency health order -- is effectively  like trying to stop an epidemic after it has become viral in that with this kind of instantaneous stopping condition, under simple assumptions, such a regulation is too little, too late. On the other hand, a graduated response that scales with increasing violence can drive equilibrium gun prevalence to zero. This is a new mathematical model relating gun violence with gun prevalence which we hope continues to spur attention from the modeling community to  bring its tools and techniques to bear on this important and challenging social problem. More importantly, our model, despite its simplicity, exhibits complex dynamics and has substantial research and educational value in promoting the application of mathematics in the social sciences, extending beyond gun control issues.

\end{abstract}


\end{frontmatter}


\section{Introduction}
\label{intro}

Laws regulating the carrying of firearms in public in the United States vary from state-to-state. As of September 2, 2023 twenty-seven of the fifty states allow for so-called ``permitless carry" \cite{PermitlessCarry}.
\footnote{``Permitless" means that there is no ``extra" permit required for carrying a firearm in public. Even in states with ``Permitless carry" there are various age requirements as well as at times other legal restrictions. See www.handgunlaw.us for details.} As a reaction to a spate of gun-related violence in New Mexico, on September 8, 2023 the Governor of New Mexico issued an emergency health order temporarily banning the carrying of firearms in public in any city or county as well as a general ban on carrying when on state property~\cite{NMHealthOrder}. More specifically, the emergency order declared a threshold of violent crime (1000 incidents per 100,000 residents) as well as  firearm-related visits to hospital emergency rooms (90 per 100,000 residents), as the marker for a 30-day suspension of the current rights of any citizen to carry a firearm. 

The action suggests a mental model of an increased likelihood of violent encounter when firearms are present, as well as violent outcomes.  As per the context of the order -- as an emergency health order -- the Governor's action bears great similarity to actions taken during pandemics, making efforts to reduce transmission of disease by reducing the likelihood of contact between the infected and susceptible populations. In the case of biological health such efforts are supported by variations on the well-studied mathematical ``SIR models'' of epidemiology. In particular, the COVID-19 pandemic inspired new models that include various forms of intervention (see e.g., Ref.~\cite{mondal2022}).

 Despite the widely proclaimed and publicized ``epidemic" of gun violence in the United States (see e.g., Ref.~\cite{politico}), that is often linked to its  outsized (as compared with the rest of the world) prevalence of gun ownership (according to World Population Review~\cite{WPR}, as of 2023 the United States could claim 120 firearms per 100 people, a rate twice as high as its nearest competitor, The Falkland Islands, both much smaller and lacking populous cities), unlike in the case of biological epidemics there has been a dearth of mathematical modeling related to the phenomenon, with most analyses focusing on correlations (positive and significant) between gun ownership rates in the United States and various harmful outcomes. Of note is the important work of Nodarz and Komarova~\cite{Wodarz2013} who consider a suite of models related to armed encounters and in particular, bring to bear techniques from optimization theory to find optimal rates of gun ownership (encoded in a ``gun control" parameter) to minimize (taking into account other parameters including gun availability, the character of the ``victim population) the death rate in the scenarios of ``one-against-one" and ``one-against-many" attacks. Their clear articulation of assumptions and the simplicity of their models is an excellent starting point for this kind of important work. More generally, understanding crime through the use of mathematical modeling, such as evolutionary games~\cite{perc2013understanding}, represents an area where computational modeling and a complex systems approach can be leveraged to address pressing social challenges~\cite{d2015statistical,jusup2022social}.

 In this paper we present a simple dynamical systems model for firearm-carrying that incorporates a ``braking term" reflecting the effect of an intervention  that is linked to measureable threshold of assumed firearm-related violence. This is consistent with the actions in the New Mexico Governor's emergency order which implicitly links firearm availability to societal violence, an assumption supported by recent rigorous studies of the effects of right-to-carry laws (cf.~\cite{Donohue2017,Donohue2019} and the references therein).

\section{Model}

Here we consider the dynamics governing the prevalence of daily firearm-carry in a region of interest, with the implicit assumption  that a greater incidence of carrying will mean more violence (as supported by Refs.~\cite{Donohue2017} and~\cite{Donohue2019}). In the absence of any restrictions, we assume that the prevalence of guns (arming rate per capita), $x$, grows as individuals choose to be armed with a rate $a$ until reaching a saturation level $k$. The parameter $0<k \le 1$ may be linked to many underlying factors including the prevailing gun culture and perception of community safety (as measured by the crime rate) in that region~\cite{Pierre2019}.  There are many reasons that people own guns: hunting (and other forms of recreation) and self-defense are the most notable open reasons, although there are also in some cases subtle psychological effects at play~\cite{Buttrick2020}. 

Logistic growth dynamics provide a plausible model for the diffusion of gun ownership:
\begin{equation}
\frac{dx}{dt} = a x (k - x).
\end{equation}
Note that this elides the division of the population into those who will carry legally and those who will carry illegally (or whom might or would violate any restrictions put in place on their carrying in public).

Let $I_g$ denote the rate of gun-related violence. We use the nonlinear equation $I_g(x) = c x^{\theta}$ to represent the  relationship between gun-owning and gun-related violence~ (see also Ref.~\cite{reeping2019state}). The constant scaling parameter $c$ also determines the ceiling rate of gun-related violence, and the parameter $\theta$ determines the curvature,  encoding the increase in gun prevalence $x$ as  it impacts gun violence:  for $\theta > 1$, we have a ``fast saturating scenario" in which gun violence $I_g$ is sensitive to $x$ and even a low gun prevalence can cause disproportionately high gun violence;  when $\theta =1$ we have a linear relation while $\theta < 1$ reflects a slowly accelerating scenario. 

In the wake of pervasive gun violence, local authorities may mimic the New Mexico Governor and respond with an intervention order aimed at suppressing the prevalence of guns within their administrative region. We assume that this intervention is linked to some statistic $\tau$. For example, New Mexico Governor Grisham uses a critical threshold of $\tau$ based on the rates of violent crime and gun-related hospital emergency room visits per 100,000 residents, to declare carrying illegal. Without loss of generality, we assume the intervention response $c(x)$ of local authorities to curb gun violence follows a sigmoid function:
\begin{equation}
c(x) = b \frac{I_g(x)^n}{\tau^n +I_g(x)^n }.
\label{sigm}
\end{equation}
The parameter $b$ quantifies the maximum strength of the intervention and the parameter $n$ controls the shape of the sigmoid function (see Fig.~1). With increasing $n$, the response approaches the limit of a step function.   

\begin{figure}[t!]
  \centering
   \includegraphics[width=0.8\columnwidth]{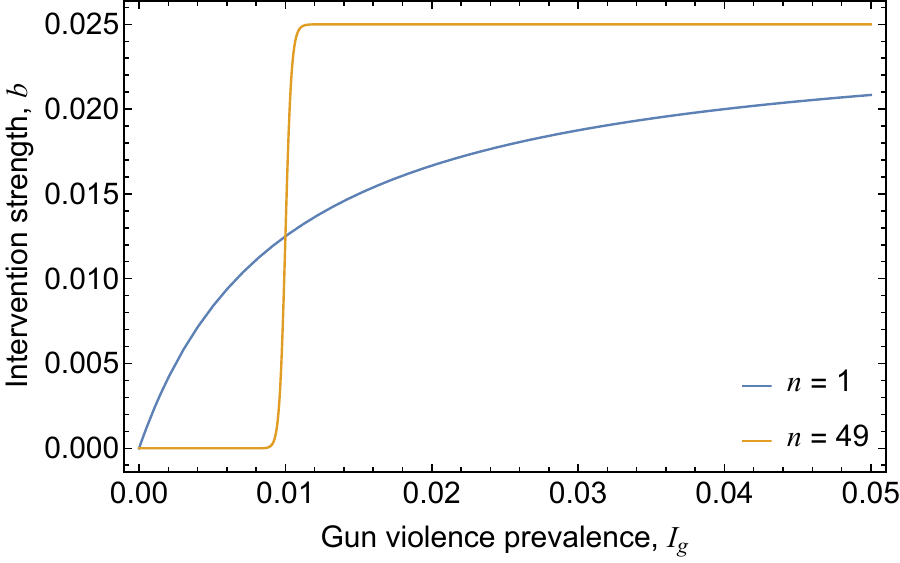} 
   \caption{Government intervention response to curb gun violence can be driven by a certain threshold in rates of gun-related violent crime. We use a sigmoid function to model threshold-dependent intervention response as described in Eq.~\ref{sigm}. The threshold $\tau = 1000/100,000$, $b = 0.025$, $n = 1$ and $n = 49$. }
   \label{fig1}
\end{figure}

In the presence of intervention order $c(x)$, the prevalence of guns is determined by the following differential equation:
\begin{equation}
\frac{dx}{dt} = a x (k - x) - c(x).
\label{gs}
\end{equation}

\section{Results}
The nonlinear dynamics described by Eq.~\ref{gs} allows bistability for most parameter choices. This means that gun prevalence can persist at high levels even with significant efforts to curb it. This is possible since as per the remark above,  some will carry in some (or all) circumstances regardless of what is allowed by law.  To demonstrate, we show a bifurcation diagram of the equilibrium prevalence rate $x^*$ with respect to varying the intervention strength parameter $b$ (see Fig.~2). In this example, we mimic the real-world scenario recently encountered in New Mexico and use a step-function like response (by setting $n = 49$ as shown in Fig.~1) that is determined by the threshold $\tau =1000/100,000$ that depends only on the violent crime rate. We also set $\theta=1$. Other choices will not qualitatively change the figure. 

\begin{figure}
   \centering
   \includegraphics[width=0.8\columnwidth]{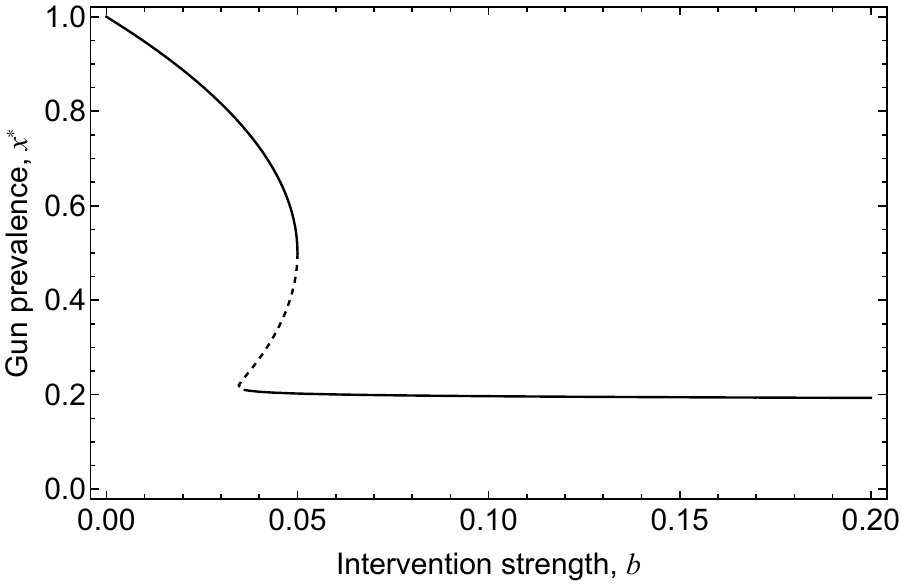} 
   \caption{Gun prevalence can persist even under step-function-like intervention responses, such as the on-and-off implementation of executive orders. This figure demonstrates that, in this model, gun violence cannot be halted by sudden, threshold-driven intervention responses, no matter how strong they are. Shown is the bifurcation diagram of equilibrium gun prevalence $x^*$ with respect to intervention strength $b$ (solid line corresponds to stable equilibrium while dashed unstable). Parameters: $a = 0.2$, $k = 1$, $c = 5000/100,000$, $\theta = 1$, $\tau =1000/100,000$ and $n = 49$. }
   \label{fig2}
\end{figure}

On the other hand, if intervention response is less abruptly driven by a prescribed threshold, but instead steadily increases with the gun violence rate, it is possible to eradicate gun violence. This could be effected in something of a step-wise fashion by requiring training, permitting, or other forms of restrictive access to the right to carry (which is different from the right to own) firearms.

To contrast with Fig.~2, we show another example in Fig.~3  setting the response curvature parameter $n = 1$ (as used in Fig.~1). In this scenario, gun prevalence can be driven to zero as long as the intervention strength $b$ exceeds a critical threshold $b_c$, ($b_c \approx 0.07$ in Fig.~3).  
\begin{figure}
   \centering
   \includegraphics[width=0.8\columnwidth]{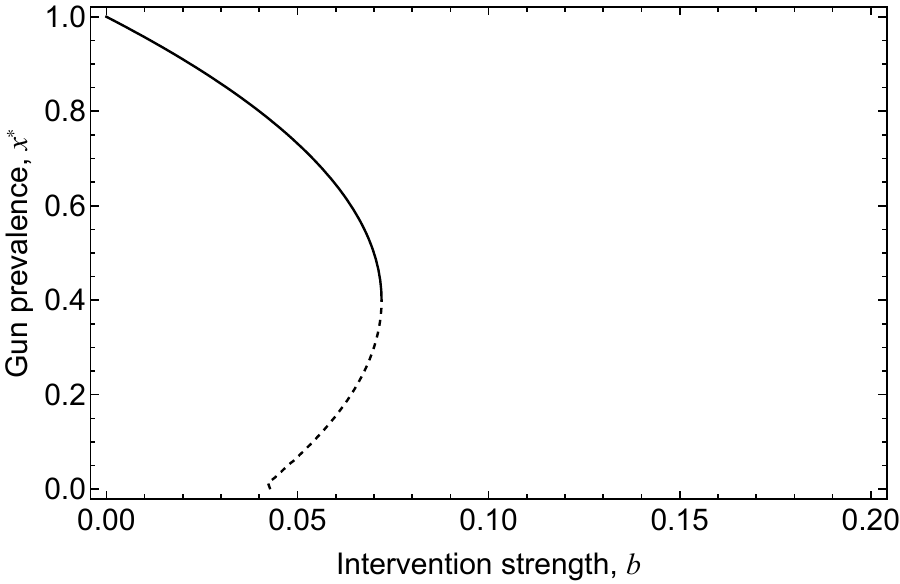} 
   \caption{The suppression of gun prevalence and, consequently, gun violence requires a steady intervention response that aims at deterring gun carrying before we arrive at a time where a dramatic intervention is effectively ``too little and too late" to control the outgrowth of gun-related violent crime. This figure shows that gun violence can be curbed with a steady response as long as the intervention strength exceeds a certain threshold, $b_c$. Shown is the bifurcation diagram of the equilibrium gun prevalence $x^*$ with respect to the intervention strength $b$ (solid line corresponds to stable equilibrium while dashed unstable). Parameters: $a = 0.2$, $k = 1$, $c = 5000/100,000$, $\theta = 1$, $\tau =1000/100,000$ and $n = 1$.}
   \label{fig3}
\end{figure}

\newpage
\section{Discussion}
Herein we present a simple dynamical systems model for firearm prevalence as a proxy for carrying firearms in public (concealed or open) with a logistic growth term impeded by an intervention term, as a sigmoid that depends on a function related to the level of violence, assumed to be measured in some fashion. Recent studies positively correlate violent crime with lax right-to-carry laws~\cite{Donohue2019} and provide some evidence for the spirit of interventions like that of New Mexico Governor Grisham's threshold-based response to high levels of violent crime in the Albuquerque area~\cite{NMHealthOrder}. Logistic growth is a traditional epidemiologically-inspired growth dynamic. In this setting it would be supported as characterizing a proportional reaction of the populace to a growing awareness of firearms in the region (e.g., through an awareness of more violent crime), some of that due to the public effects of right-to-carry. 

We have approached the intervention aimed at suppressing gun prevalence and violence mainly from a top-down perspective, namely, the removal of gun carriers at a rate that depends on the strength of intervention. This is of course an oversimplification. Extensions of our current model to account for individual perception and compliance of the intervention are meaningful and necessary next steps for a fine-grained modeling-based understanding. In particular, prior work on adaptive compliance of social distancing measures suggests that an oscillating tragedy of the commons is likely in the context of gun-related violence and interventions~\cite{glaubitz2020oscillatory}.    

In our model a sharp intervention produces a nonzero equilibrium firearms prevalence, while a modulated response that scales as violence increases produces a stable equilibrium of zero prevalence. The latter could be effected in part by levels of restricted access to the right-to-carry in public. We hope that this model or dynamical systems model like this (also see Ref.~\cite{monteiro2020more}) might provide food-for-thought for policymakers as well as future efforts to assist in battling the problem of gun violence in the United States. 

\section*{Acknowledgments}
F.F. gratefully acknowledges support from the Bill \& Melinda Gates Foundation (award no. OPP1217336).

\section*{Code Availability}
The code to reproduce the reported results is publicly available at \url{https://github.com/fufeng/gunviolence/}.

\section*{Author Contributions Statement}
The two authors contributed equally to all aspects of the paper.


\end{document}